\documentclass[10pt, letterpaper]{emulateapj}

\begin{document}
 
\title{Interferometric Observations of V1663 Aquilae (Nova Aql 2005)}
\author{B. F. Lane \altaffilmark{1}, A. Retter \altaffilmark{2}, J. A. Eisner\altaffilmark{5},
M. W. Muterspaugh\altaffilmark{3}, R. R. Thompson \altaffilmark{4},
J. L. Sokoloski\altaffilmark{6}}

\altaffiltext{1}{Kavli Institute, MIT Department of Physics, 70 Vassar Street, Cambridge, MA 02139; blane@mit.edu}
\altaffiltext{2}{Astronomy \& Astrophysics Dept., Penn State University, 525 Davey Lab, University Park, PA 16802-6305; retter@astro.psu.edu}
\altaffiltext{3}{Townes Fellow, Space Sciences Laboratory, University of California, Berkeley, CA 94720}
\altaffiltext{4}{ Palomar Testbed Interferometer, Palomar Mtn, CA 92060}
\altaffiltext{5}{Miller Fellow, University of California at Berkeley, 601 Campbell Hall, Berkeley, CA 94720}
\altaffiltext{6}{Columbia Astrophysics Lab, MC 5247, Columbia University, New York, NY 10027.}
\begin{abstract} 
We have resolved the classical nova V1663 Aql  using long-baseline
near-IR interferometry  covering the period from $\sim$5--18 days after 
peak brightness. We directly measure the shape and size
of the fireball, which we find to be asymmetric. In addition 
we measure an apparent expansion rate of  $0.21 \pm
0.03$ ${\rm mas\,day^{-1}}$. Assuming a linear expansion 
model we infer a time of initial outburst approximately 4 days prior to peak brightness. 
When combined with published spectroscopic expansion velocities our 
angular expansion rate implies a distance of
$8.9\pm3.6$ kpc. This distance measurement is independent of, but
consistent with, determinations made using widely available
photometric relations for novae. 
\end{abstract}


\section{Introduction}
Novae are violent stellar explosions exceeded in energy output only by
$\gamma$-ray bursts and supernovae. They are erratic outbursts that
occur in systems containing a white dwarf accreting mass from a
late-type stellar companion (e.g Prialnik \& Kovetz\nocite{pk05} 2005). 
When the amount of accreted
material on the surface of the white dwarf reaches some critical value
a thermonuclear-runaway is ignited, giving rise to the observed nova
outburst in which material enriched in heavy elements is ejected into
the surrounding medium at high velocities. For certain elements, this
ejected material may influence observed abundances in the
ISM \citep{gehrz98,hernanz05}. Direct observations of the expansion of
the nova shell provide an opportunity to accurately determine the
distance to the nova. Such observations are usually only possible many
months or years after the outburst, when the expanding shell can be resolved
\citep{bode02}. However, several optical/IR interferometers are now capable of resolving 
novae and other explosive variables, allowing detailed studies of the 
initial "fireball", \citep{ches07,mon06}  as well as later stages of 
development \citep{lane05,lane07}.

Nova Aquilae 2005 (ASAS190512+0514.2, V1663 Aql) was discovered on 9
June 2005 by \citet{iauc8540}.  At the time of discovery the magnitude
was $m_V$ = 11.05; the source reached $m_V \sim 10.8$ the following
day, and declined in brightness thereafter. A possible progenitor near
the source coordinates (sep. $\sim$ 4.5 arcseconds) is seen on Palomar
Optical Sky Survey plates (USNO-B1.0 0952-00410569, \citet{usno-b1}),
with magnitudes $m_R \sim 18.1$ and $m_I \sim 16.45$.  Soon after
discovery \citet{iauc8544} obtained an optical spectrum with features
indicating a heavily reddened, peculiar nova. H-$\alpha$ emission
lines exhibited P Cygni line profiles and indicated an expansion
velocity of $700 \pm 150 {\rm km\,s^{-1}}$ \nocite{iauc8544} (Dennefeld et al. 2005, 
confirmed via personal communication), somehwhat slow for a classical nova, but not outside
the range of observed values. Recently, \citet{pog06} published
spectra and analysis of published light-curves of this nova, deriving
a distance in the range 7.3--11.3 kpc, and an expansion velocity of $\sim2000$~${\rm km\,s^{-1}}$.

We have used the Palomar Testbed Interferometer (PTI) to resolve the
$2.2 \mu$m emission from V1663 Aql and measure its apparent angular
diameter as a function of time. We are able to follow the expansion
starting $\sim 9$ days after the initial explosion; when combined with
radial velocities derived from spectroscopy we are able to infer a
distance to, and luminosity of, the object. We compare this result
with values inferred by a maximum magnitude-rate of decline (MMRD)
relation in the literature (see Poggiani 2006 for a summary).

The Palomar Testbed Interferometer
(PTI) was built by NASA/JPL as a testbed for developing ground and
space-based interferometry and is located on Palomar Mountain near San
Diego, CA \citep{col99}. It combines starlight from two out of three
available 40-cm apertures and measures the resulting interference
fringes. The high angular resolution provided by this long-baseline
(85-110 m), near infrared ($2.2 \mu$m) interferometer is sufficient to
resolve emission on the milli-arcsecond scale.

\section{Observations}

We observed V1663 Aql on 10 nights between 15 June 2005 and 28 June
2005; on six of those nights we obtained data on two or three
interferometric baselines.  For a detailed description of the
instrument we refer the reader to \cite{col99}.  Each nightly
observation consisted of one or more 130-second integrations during
which the normalized fringe visibility of the science target was
measured. The measured fringe visibilities of the science target were
calibrated by dividing them by the point-source response of the
instrument, determined by interleaving observations of calibration
sources (Table \ref{tab:calibs}); the calibration sources were chosen
to be single stars, close to the target on the sky and to have angular
diameters less than 2 milli-arcseconds, determined by fitting a
black-body to archival broadband photometry of the sources.  For
further details of the data-reduction process, see \citet{colavita99b}
and \citet{boden00}. Calibrated fringe visibilities are listed in
Table \ref{tab:data}; though given as averages, the fits to the models
were done using the individual data points, without averaging on a nightly basis.
Note that HD~188310 has been claimed to be a binary 
with a separation of $0.09\pm0.01$ arcseconds \citep{scardi00}. However, 
this claim is based on a single observational epoch obtained under poor 
observing conditions, and has to our knowledge not been confirmed
by any other group. We do not see any indications of binarity in 
our data; the fringe visibilities of HD~188310 are stable to 2.5\%
when calibrated using HD~176303. In addition, we reduce the V1663 
Aql data both with and without using HD~188310 as a calibrator and find the 
results to be fully self-consistent. We include HD~188310 as a calibrator in
the results shown here as it was the only calibrator available on the last 
night of observations. 

\begin{table}
\begin{center}
\begin{tabular}{lccccc}
           &       &   &             &                     &  \\
\tableline
\tableline
Calibrator & V     & K & Spectral    & Ang. Diameter       &    Separation \\
           &       &   & Type        & $\theta_{UD}$ (mas) &     (deg) \\
\tableline
HD~176303  & 5.27   & 3.93 & F8V       & $0.71 \pm 0.03$     & 8.5 \\
HD~188310  & 4.72   & 2.17 & G9III     & $1.62 \pm 0.08$     & 13 \\
HD~186442  & 6.56   & 3.67 & K0III     & $0.92 \pm 0.16$     & 11 \\
\tableline
\end{tabular}
\caption{\label{tab:calibs} Relevant parameters of the calibrators and check star. The 
separation listed is the angular distance from the calibrator to V1663 Aql.  Alternate catalog designations: HD~176303 = 11 Aql = HR 7172; HD~188310 = 59 Aql = $\xi$ Aql = HR 7595; HD 186442 = BD +09 4233 = SAO 125000.}
\end{center}
\end{table}

\begin{table}
\begin{center}
\begin{tabular}{lcccc}
        &      &      &          &                \\
\tableline
\tableline
 Epoch  &  $u$ &  $v$ & No. Pts. &Cal. Visibility \\
 (MJD)  &  (m) &  (m) &          & ($V^2$)        \\
\tableline
53536.424 & -75.3240 & -25.0938 &  1 &$    0.698 \pm 0.117$  \\ 
53537.393 & -79.8978 & -23.7842 &  8 &$    0.749 \pm 0.030$  \\ 
53538.383 & -44.1836 &  66.0275 & 10 &$    0.718 \pm 0.033$  \\ 
53539.346 & -81.4086 & -21.8080 & 14 &$    0.672 \pm 0.050$  \\ 
53540.392 & -79.3017 & -24.1591 &  4 &$    0.610 \pm 0.035$  \\ 
53540.333 & -51.6711 & -88.3761 &  5 &$    0.567 \pm 0.026$  \\ 
53541.395 & -78.3316 & -24.4178 &  3 &$    0.568 \pm 0.038$  \\ 
53541.339 & -48.6701 & -88.6301 &  5 &$    0.484 \pm 0.024$  \\ 
53542.396 & -77.4187 & -24.5759 &  3 &$    0.538 \pm 0.037$  \\ 
53542.320 & -53.4276 & -88.0947 &  6 &$    0.534 \pm 0.048$  \\ 
53542.458 & -59.4251 &  63.4296 &  5 &$    0.519 \pm 0.074$  \\ 
53543.400 & -76.2535 & -24.8719 &  4 &$    0.499 \pm 0.029$  \\ 
53543.452 & -59.1619 &  63.5118 &  5 &$    0.454 \pm 0.046$  \\ 
53548.363 & -80.2661 & -23.8083 &  5 &$    0.537 \pm 0.078$  \\ 
53548.300 & -54.5455 & -88.0372 &  5 &$    0.539 \pm 0.096$  \\ 
53549.377 & -77.8030 & -24.5517 &  2 &$    0.614 \pm 0.194$  \\ 
53549.305 & -52.5498 & -88.2695 &  4 &$    0.538 \pm 0.173$  \\ 
\tableline
\end{tabular}
\caption{\label{tab:data} Calibrated fringe visibilities of
V1663 Aql, together with the projected baseline components.
The effective wavelength of observations was 2.2 $\mu$m.
 $u,v$ are the components of the projected baseline in meters. All of the 
data from a given baseline on a given night has been averaged. }
\end{center}
\end{table}

PTI is equipped with a low-resolution spectrometer which provides
fringe visibility measurements and photon count rates in five spectral
channels across the K band. In addition to the fringe visibilities
measured in each channel (referred to as ``narrow-band'' visibilies),
we compute a wide-band average visibility as the photon-weighted
average of the five spectral channels.  Using photon count rates from
PTI and K-band magnitudes for the calibrator sources provided by 2MASS
\citep{2mass}, we also derive a K-band apparent magnitude of V1663
Aql.  It should be remembered that PTI was not designed 
for high-precision photometry, and hence the K-magnitudes, while useful, should be
treated with some caution.

 \section{Models \& Results}

\subsection{Interferometric Data}
 
The theoretical relation between source brightness distribution and
fringe visibility is given by the van Cittert-Zernike theorem\citep{tms01}.
For a uniform intensity disk model the normalized fringe visibility (squared) can be related
to the apparent angular diameter using
\begin{equation}
V^2 = \left( \frac{2 \; J_{1}(\pi B \theta_{UD} / \lambda)}{\pi B\theta_{UD} / \lambda} \right)^2
\label{eq:V2_single}
\end{equation}
where $J_{1}$ is the first-order Bessel function, $B$ is the projected
aperture separation and given by $B = \sqrt{u^2 + v^2}$ where $u,v$
are two orthogonal components of the projected baseline, $\theta_{UD}$
is the apparent angular diameter of the star in the uniform-disk
model, and $\lambda$ is the wavelength of the observation. 

\begin{figure}[h]
\epsscale{1.0}
\plotone{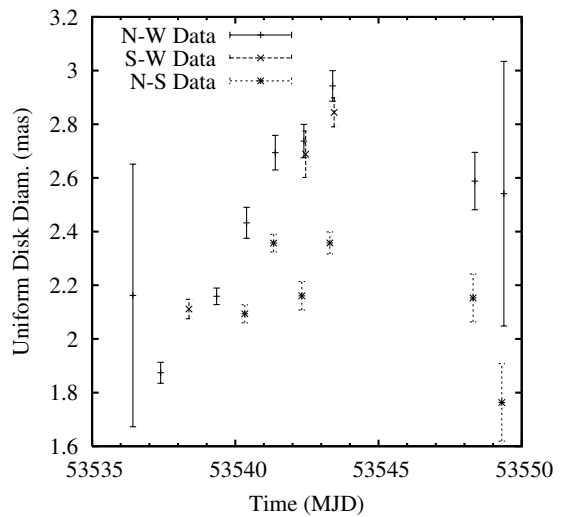}
\caption[]{\label{fig:udfits} The apparent angular dimeter of the 
V1663 Aql source, as a function of time, assuming a uniform (circular) 
disk model. Each symbol represents a fit to data from a single baseline.
Although the data are all consistent with an expansion, the disagreement 
between various baselines indicates the need for a more complex
morphological model.}
\end{figure}

We obtain an initial characterization of the data using a circularly symmetric 
uniform-disk model fit to data from a single baseline on a nightly basis (Fig. \ref{fig:udfits}).
This indicates that the apparent source morphology was initially 
expanding, but that the apparent expansion reversed sometime between
MJD 53545 and MJD 53548. The data also clearly indicate that while
the North-West and South-West data produce consistent angular diameters, the 
North-South data indicates both a smaller apparent diameter and a 
less smooth expansion. Clearly, a more sophisticated morphological model 
is required. 

The availability of data taken with multiple baselines with different
position angles allows us to distinguish between a circularly
symmetric source and elliptical  disk models.  For the
asymmetric cases we fit for three parameters: size ($\theta$),
inclination angle ($\phi$), and position angle ($\psi$). Inclination
is defined such that a face-on disk has $\phi=0$ and $\psi$ is
measured east of north. Following \citet{eis03}, we include $\phi$ and
$\psi$ in our models of the brightness distribution via a simple
coordinate transformation:
\begin{eqnarray}
u\prime & = & u \sin \psi + v \cos \psi \\
v\prime & = & \cos \phi \left ( v \sin \psi - u \cos \psi \right ) 
\end{eqnarray}
Substitution of ($u\prime, v\prime$) for ($u, v$) in the expressions
above yields models with inclination effects included. Note that the
``inclination'' is not tied to a true disk-like morphology; the source
can well be (and in this particular case probably is) an ellipsoid,
in which case the major axis of the ellipsoid is $\theta$ and the 
minor axis is given by $\theta \cos(\phi)$. We perform least-squares
fits of uniform and inclined disk (or equivalently ellipsoidal) models
to the measured fringe visibilities.

\begin{figure*}[htbp]
\epsscale{1.0}
\plottwo{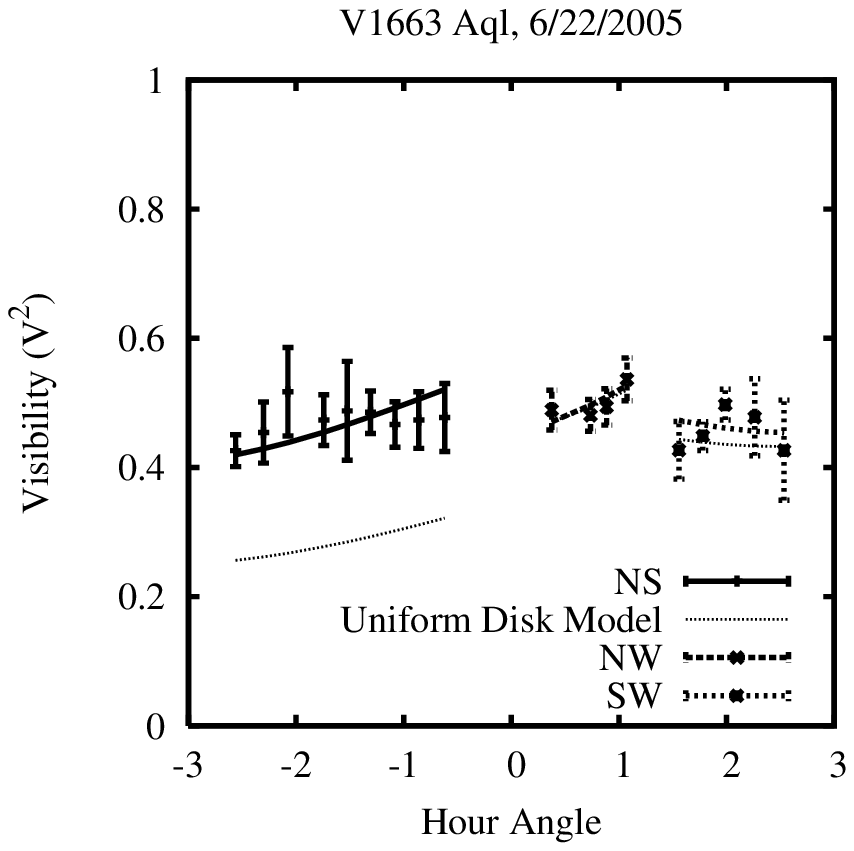}{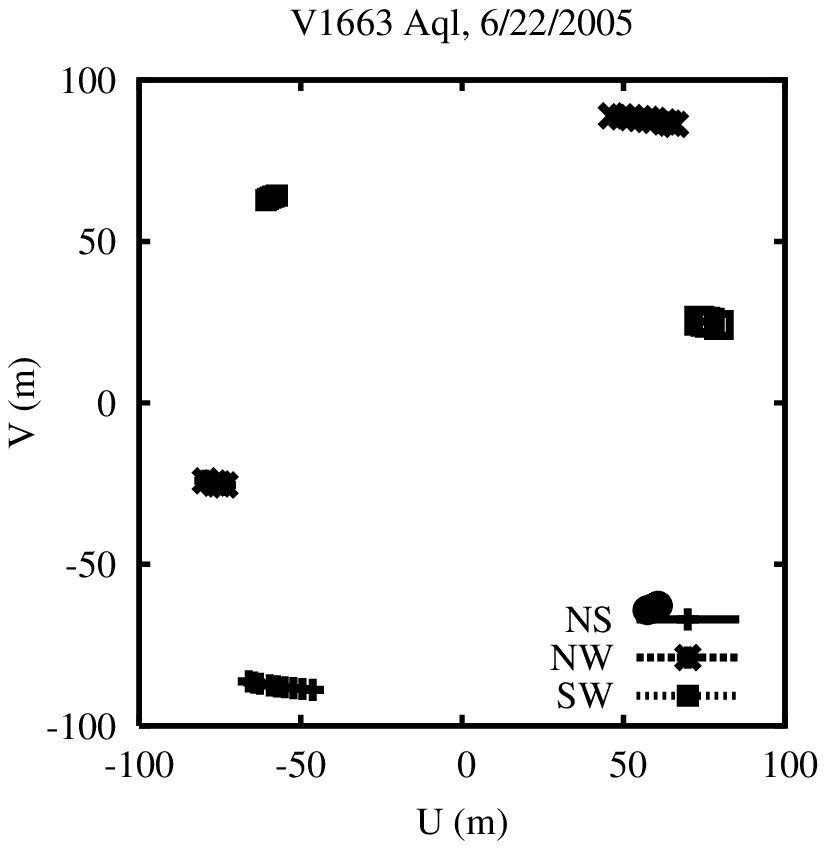}
\caption[]{\label{fig:inclined} (left) The measured fringe visibility of
V1663 Aql as a function of hour angle during the night of 22 June
2005. Although PTI can only use one baseline at a time, during the
night two changeovers were performed, yielding data on all three
baselines within a relatively short period of time. An inclined-disk
model was fit to the data (shown as thick lines), yielding a good fit. The 
fit parameters are given in Table \ref{tab:dfits}. A corresponding 
uniform-disk fit (thin lines) is shown for comparison; while it can match
the N-W and S-W data it significantly under-predicts the N-S baseline 
visibilities. (right) The $uv$-plane coverage (i.e. the projected baselines)
of the observations.}
\end{figure*}

As is apparent from Fig. \ref{fig:inclined} the data on nights with
more than one available baseline are not well matched by a simple
uniform disk model; in effect the visibility in the North-South
baseline is too high (and the corresponding uniform-disk diameter too
small) to be consistent with the North-West and South-West
baselines. Typical goodness-of-fit parameters are
$\chi_r^2\sim5$. However, it is possible to fit the data using an
inclined disk-model (Table \ref{tab:dfits}). While the angular size
scale changes from night to night, the consistency of the inclination
and position angles of the fits lend credence to the idea that we are
observing an expanding, asymmetric source.

\begin{table*}
\begin{center}
\begin{tabular}{lccccc}
                    &            &                       &                &             \\
\tableline
\tableline
Date                 &  $\chi_r $&   Size                & $\psi$          & $\phi$   & Baselines\\
                     &             &   (mas)               &  (deg)          &  (deg)   &      \\
\tableline
19 June 2005 &  0.4    &  $3.1\pm0.5$      & $129.5 \pm 6.6$  & $49.2\pm8.3$  & NS,NW\\
20 June 2005 &  0.8    &  $3.3\pm0.5$      & $127.7 \pm 7.9$  & $46.0\pm9.2$  & NS,NW \\
21 June 2005 &  0.5    &  $2.9\pm0.05$    & $ 103.8 \pm 2.9$ & $44.8\pm2.0$ & NS,NW,SW \\
22 June 2005 &  0.6    &  $3.1\pm0.05$    & $101.6 \pm 2.1$ & $45.11\pm2.0$ & NS,NW,SW \\
27 June 2005 &  0.4    &  $3.2\pm1.0$       & $ 127.1 \pm 19$  & $48.8\pm18$ & NS,NW \\
28 June 2005 &  4.4    &  $2.8\pm1.7$       & $ 80.7 \pm 161$  & $89.2\pm360$ & NS,NW \\

\tableline
\end{tabular}
\caption{\label{tab:dfits} Inclined-disk model fits to data from a given
multi-baseline night. NS = North-South, NW = North-West and 
SW  South-West baselines, respectively}
\end{center}
\end{table*}

We examine the possibility of systematic errors mimicing the effect of
non disk-like emission using the following test: we use HD 186442 as a
``check star'' by making observations of it interleaved with
observations of our calibrators and V1663 Aql. Such observations were
made on three nights, and on two of those nights we observed the check
star with all three available interferometric baselines.  We calibrate
the check star data using the same procedure as applied to V1663 Aql, and
fit a uniform-disk model to the calibrated nightly data. We find that
the best-fit uniform disk diameter is $0.83 \pm 0.05$ milli-arcseconds
with a goodness-of-fit parameter ($\chi_r$) of 0.80 for 16 130-second
data points, and a mean scatter in the visibilities about the best-fit
model of 2.9\%, fully comparable with typical point source observations 
using PTI. 

\begin{figure}[h]
\epsscale{1.0}
\plotone{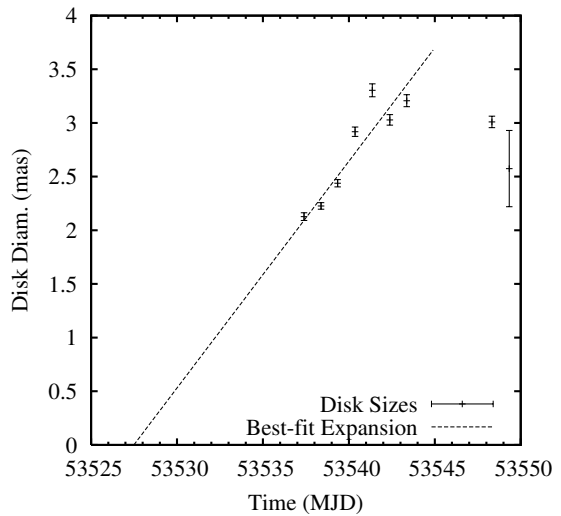}
\caption[]{\label{fig:expansion} The best-fit inclined disk diameter as 
a function of time, togther with a best-fit linear expansion model. 
The expansion appears to stop sometime after MJD 53545; the subsequent
points have been left out of the fit. }
\end{figure}

Given the reasonable stability of the position angle and inclination (or
equivalently, aspect ratio) parameters, we re-fit the data from all
the nights, holding these two parameters fixed at the average values
from Table \ref{tab:dfits} (excluding the last two points where 
the fits are effectively indeterminate), while
letting the diameter vary on a nightly basis. The resulting best-fit
diameters are given in Table \ref{tab:fits} and shown in Figure
\ref{fig:expansion}.  A simple linear expansion model fit to the disk
sizes indicates an expansion rate of $0.21 \pm 0.03$
milli-arcseconds/day, beginning on MJD $53527.4 \pm 1.9 $.

\begin{table}
\begin{center}
\begin{tabular}{lccccc}
                &            &             &               \\
\tableline
\tableline
Date            & MJD        &    Size        &  $\chi_r $ \\      
                &            &    (mas)       &              \\     
\tableline
 16 June 2005 &53537.393 & $2.13 \pm 0.03$& 0.7\\
 17 June 2005 &53538.383 & $2.23 \pm 0.03$& 0.6\\
 18 June 2005 &53539.346 & $2.44\pm 0.03$& 0.9\\
 19 June 2005 &53540.359 & $2.92 \pm 0.04$& 1.3\\
 20 June 2005 &53541.36 & $3.30\pm 0.06$& 2.3\\
 21 June 2005 &53542.385 & $3.03 \pm 0.05$& 1.2\\
 22 June 2005 &53543.366 & $3.21\pm 0.06$& 2.6\\
 27 June 2005 &53548.325 & $3.01\pm 0.05$& 0.4\\
 28 June 2005 &53549.329 & $2.57 \pm 0.35$& 3.3\\

\tableline
\end{tabular}
\caption{\label{tab:fits} The best-fit major-axis angular sizes of 
inclined-disk models fit to the data on a nightly basis. The inclination
of the disks was fixed at 46 degrees, and the position angle 
was held at 115 degrees, corresponding to the average values 
determined from the best fits in Table \ref{tab:dfits}.  }
\end{center}
\end{table}

\subsection{Narrow-band Interferometric Data}

\begin{figure}[h]
\epsscale{1.0}
\plotone{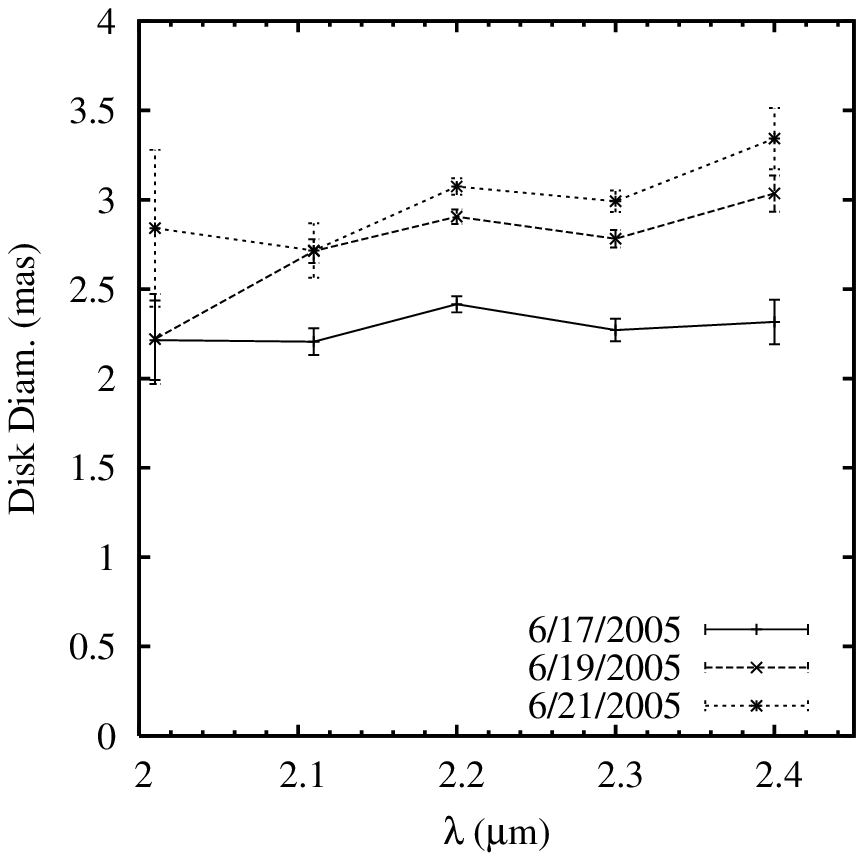}
\caption[]{\label{fig:nb} The best-fit inclined disk diameter as 
a function of wavelength, for data from three nights.  The 
inclination and posotion angles were held fixed at the 
average best-fit values from Table \ref{tab:dfits}}
\end{figure}

In addition to the wide-band fringe visibilities, we can compute the 
best-fit disk diameters for data from each spectral channel separately; 
results from three nights are show in Fig. \ref{fig:nb}.  The lack of 
significant variation in apparent angular diameter with wavelength 
supports  our assumption that the observed 
visible emission lines are originating from the same region as the K-band 
emission seen by PTI, at least during the first few weeks of the expansion.

\subsection{Photometry}
\label{sec:phot}
\begin{figure}[h]
\epsscale{1.0}
\plotone{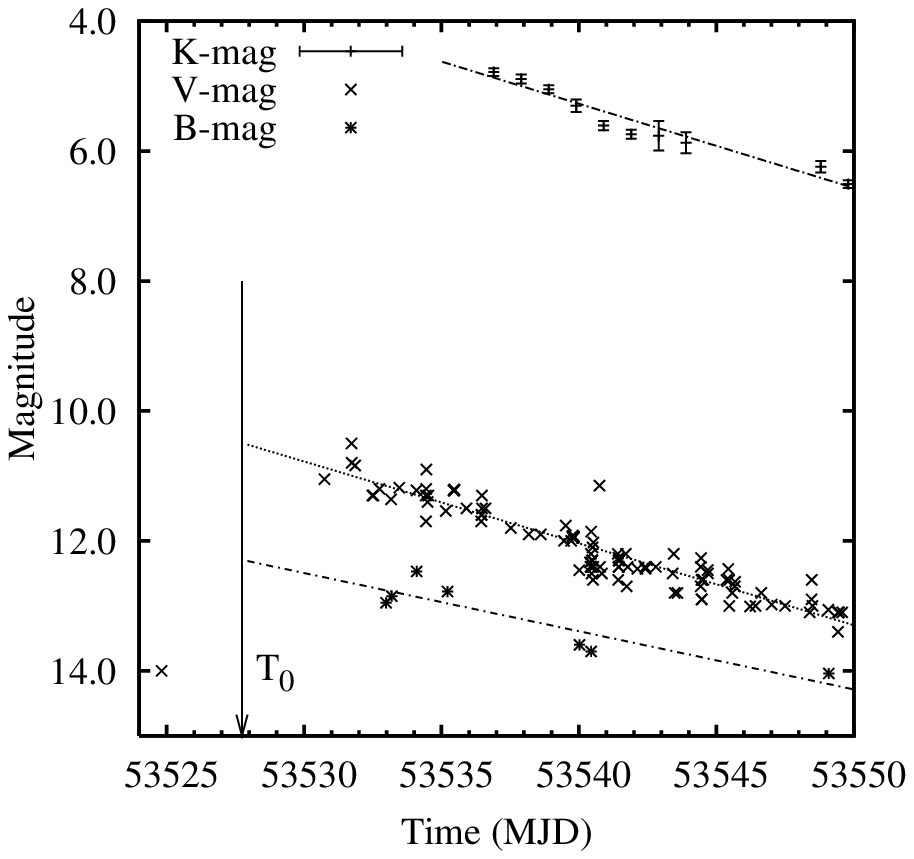}
\caption[]{\label{fig:phot} Measured B-,V- and K-magnitudes
of V1663 aql following the nova outburst. K magnitudes were 
measured by PTI, while B and V magnitudes were collected
from VSNET and IAU circulars. The best-fit linear magnitude trends
are also shown, together with the time of initial expansion
$T_0$ determined from fitting to the measured diameters.}
\end{figure}

We derive the approximate K-band magnitude of V1663 Aql using photon
fluxes measured by PTI; the results are shown in
Fig. \ref{fig:phot}. We have also collected B and V-band photometry
published from VSNET\footnote{\tt http://www.kusastro.kyoto-u.ac.jp/vsnet/} 
and IAU circulars. We fit a linear trend to the data in order to derive the rate 
of declines in the K and V bands; we find
that the PTI K-band photometry implies $\dot{m}_K = 0.13 \pm 0.01 {\rm mag\, day^{-1}}$,
and the compiled V-band photometry gives
$\dot{m}_V = 0.126 \pm 0.004{\rm mag\, day^{-1}}$. The V-band data 
indicates a time to decline 2
magnitudes from maximum ($t_2$) of $15.9\pm0.5$ days, and a 3-magnitude
decline $t_3 = 23.8 \pm 0.8$ days. We also fit a low-order polynomial 
function to the $B$- and $V$-band photometry, allowing us to find an 
approximate $B-V$ color for $t=t_2$ of 1.11. Based on the work by \citet{vdB87} the intrinsic B--V 
color of novae two magnitudes below maximum light is $-0.02\pm0.04$, 
and hence we infer $E(B-V)=1.1$. The corresponding V-band extinction
$A_V\sim 3.4$ mag.

We use the MMRD relation 
from \citet{dV95} and our value of $t_2$ to derive $M_V = -8.31$
for V1663 Aql. Adopting a peak value of $m_{V,max}=10.8$ we
would predict a distance of $\sim 13.8$ kpc.

\section{Discussion}

Using our interferometrically measured angular diameters it is
possible to constrain three important parameters of this system:
the fireball shape, the time of initial expansion, and the distance to the system. 
However, we note that we are using relatively simplistic models, 
and our interpretations are therefore limited and subject to 
a systematic uncertainty that is difficult to quantify. Nonetheless, the 
measured fringe visibility and the way it changes with time and baseline 
strongly imply that the source is expanding and likely asymmetric. 

First, we measure a non-zero asymmetry in the expanding 
fireball as early as ten days after the initial outburst; this 
would tend to indicate that the source of the asymmetry is 
inherent to the explosion mechanism and not the result 
of interactions with the interstellar or circumstellar medium. 
As discussed in \citet{bode02} there is an apparent correlation between
nova speed class and apparent asymmetry; our measured 
asymmetry (major/minor axial ratio = 1.44, for $t_3\sim24$d) is
larger than what would be expected from that relation (1.0--1.1). 

Second, we can extrapolate backward in time to find the time of the inital
explosion; we find that it occurred around MJD $53527.4 \pm 1.9
$. Comparing this with the time of maximum V-band luminosity (MJD
$53531.7 \pm 0.1$; Poggiani 2006) we find a delay of $\sim 4 \pm 2$ days
between the onset of expansion and the maximum observed luminosity.
In this context we note the somewhat odd V-band point from June 3 ($m_V=14$)
reported in \citet{iauc8540}; this is probably not the pre-outburst 
magnitude of the nova. However, this point would then indicate the 
the outburst began sometime before 3 June, i.e. even earlier than 
our $T_0$. We have no other indications of a pre-maximum halt 
in this nova and note that such phenomena are usually associated with 
slow novae. On the other hand, \citet{k02} have observed 
a long pre-maximum halt in the rapidly evolving nova V463 Sct, indicating
that such expectations cannot be absolute.

Third, we can use our measured expansion rate of $0.21 \pm 0.03$
${\rm mas\,day^{-1}}$ together with spectroscopically determined expansion speeds to
find a geometric distance to this nova, independent of any photometric
biases. Unfortunately there appears to be little consensus as to
the spectroscopically determined expansion velocity, with available 
estimates differing by a factor of three. 
Poggiani (2006) measured the H$\alpha$ line on 31 July 2005 and finds a
Half-Width at Zero Intensity (HWZI) $\sim 2000 {\rm km\,s^{-1}}$, indicating an
expansion velocity in that range, while the O I 8446 line from the same paper 
appears more consistent with a HWZI of $\sim 1500 {\rm km\,s^{-1}}$.  
On the other hand, \citet{iauc8544} find an expansion rate of 700${\rm km\,s^{-1}}$ from 
spectroscopy obtained on 11 June 2005, while \citet{iauc8640} find a
Full-Width at Zero Intensity $\sim 2600 {\rm km\,s^{-1}}$
from spectroscopy obtained on 14 Nov 2005. 
We adopt an intermediate value of 1375 $\pm 500{\rm km\,s^{-1}}$,
being the average and standard deviation of the four values listed above. 
 In this context we note that the expected expansion rate for a nova with 
$t_3 = 23.8$ days is $\sim1000  {\rm km\,s^{-1}}$  (McLaughlin 1960). 

A second difficulty, explained by \citet{wh00}, is that for a non-spherical 
nova shell the unknown orientation of the shell in the plane of the sky 
biases any distance determination that does not account for this 
inclination effect, and in fact the proper inclination can only be determined
with spatially resolved spectroscopy.  In our case, we measure an ellipse with an apparent 
minor/major axis ratio of  $\cos(\theta)=0.69$, and do not have access to 
any information that would unambiguously constrain the inclination
of the nova shell; we therefore use the estimator recommended by 
\citet{wh00}, viz. the arithmetic mean of the apparent  major and minor 
axis expansion rates. This estimator will on average yield a result 
within a few percent of the true distance as long the orientation 
of the nova shell isn't close to 0 or 90 degrees.  We find a 
distance to this nova of $8.9\pm3.6$ kpc.
This is consistent with the range derived by Poggiani (2006), based on
a number of photometric MMRD relations 
available in the literature (7.3--11.3 kpc), but marginally inconsistent 
with the value we derive in Section \ref{sec:phot}.
We caution, however, that the expansion rate may in fact change 
with time; this could explain the range of  measured rates, as well as bias our result. 

Finally, an important feature in our data is that the apparent expansion 
ceased sometime after MJD 53545 (day 18). We note that behavior is 
similar to what was seen in RS Oph \citep{lane07}, where the apparent 
expansion appeared to reverse around day 20.  It is possible that the apparent reversal is 
not due to the transition from optically thick to optically thin emission 
from the expanding fireball \citep{gehrz88b}, but rather that the wind 
mass-loss rate (and hence effective optical depth of the source material) 
is changing.  However, the details of this process 
are not easily modeled and we will explore this further in subsequent 
papers. 

\section{Conclusion}

We have used long-baseline near-IR interferometry to resolve the
classical nova V1663 Aql, starting $\sim 9$ days after outburst.  We
measure an apparent expansion rate of $0.21 \pm 0.03$ ${\rm mas day^{-1}}$, which
can be combined with previously determined expansion velocities to
produce a distance estimate to the nova of 
$8.9\pm3.6$ kpc; the precision is 
limited by the precision of the available spectroscopic radial velocities. 
Such a large distance is consistent with the large reddening ($E(B-V) \sim
1.1$) determined from photometry, as well as with distances found from
MMRD relations. This represents only the third time a nova 
has been resolved using optical/IR interferometry, the previous cases being 
Nova V1974 Cyg 1992 \citep{q93} and more recently RS Oph \citep{mon06,lane07}.
We anticipate that further instrumental improvements, in particular
high spectral resolution interferometry such as that recently deployed
on the Keck Interferometer \citep{eis07}, may break the inclination degeneracy 
and thus yield very precise expansion distances to these interesting objects.

\acknowledgements 
We wish to acknowledge the extraordinary
observational efforts of K. Rykoski. Observations with PTI are made
possible through the efforts of the PTI Collaboration, which we
gratefully acknowledge. This research has made use of services from
the Michelson Science Center, California Institute of Technology,
http://msc.caltech.edu.  Part of the work described in this paper was
performed at the Jet Propulsion Laboratory under contract with the
National Aeronautics and Space Administration. This research has made
use of the Simbad database, operated at CDS, Strasbourg, France, and
of data products from the Two Micron All Sky Survey, which is a joint
project of the University of Massachusetts and the Infrared Processing
and Analysis Center/California Institute of Technology, funded by the
NASA and the NSF. MWM is grateful for the 
support of a Townes Fellowship.

\end{document}